\documentclass{aa}
\usepackage{graphicx}

\begin{document}
   \title{Testing the predicted mass-loss bi-stability jump at radio wavelengths}

   \author{P. Benaglia
          \inst{1,2}
          \and J. S. Vink\inst{3,4}
          \and J. Mart\'{\i}\inst{5}
          \and J. Ma\'{\i}z Apell\'aniz\inst{6,7}
          \and B. Koribalski\inst{8}
          \and P.A. Crowther\inst{9}
          }

   \offprints{P. Benaglia}

   \institute{Instituto Argentino de Radioastronom\'{\i}a, C.C.5, (1894) Villa Elisa, Argentina.
              \email{pbenaglia@fcaglp.unlp.edu.ar}
         \and Facultad de Cs. Astron\'omicas y Geof\'{\i}sicas, UNLP, Paseo del Bosque s/n,
(1900) La Plata, Argentina
          \and Keele University, Astrophysics, Lennard-Jones Lab, ST5 5BG, UK.
              \email{jsv@astro.keele.ac.uk}
          \and Imperial College, Blackett Laboratory, Prince Consort Road, London, SW7 2AZ, UK
          \and Departamento de F\'{\i}sica, EPS, Universidad de Ja\'en, Campus Las Lagunillas
s/n, Edif. A3, 23071 Ja\'en, Spain.
             \email{jmarti@ujaen.es}
          \and Instituto de Astrof\'{\i}sica de Andaluc\'{\i}a, Camino bajo de Hu\'etor 50, Granada 18008, Spain.               \email{jmaiz@iaa.es}
          \and Ram\'on y Cajal fellow, Ministerio de Educaci\'on y Ciencia, Spain.
          \and Australia Telescope National Facility, CSIRO, PO Box 76, Epping, NSW 1710, Australia.
              \email{Baerbel.Koribalski@csiro.au}
          \and  Department of Physics and Astronomy, University of Sheffield, Hicks Building, Hounsfield Road, Sheffield S3 7RH, UK.
              \email{Paul.Crowther@sheffield.ac.uk}
          }

   \date\today

  \abstract
   {Massive stars play a dominant role in the Universe, but one of the main drivers for their
    evolution, their mass loss, remains poorly understood.}
   {In this study, we test the theoretically predicted mass-loss behaviour as a function
   of stellar effective temperature across the so-called `bi-stability' jump.}
   {We observe OB supergiants in the spectral range O8-B3 at radio wavelengths to
   measure their thermal radio flux densities, and complement these measurements with data from the literature.
   We derive the radio mass-loss rates and wind efficiencies, and compare our results with H$\alpha$ mass-loss rates
   and predictions based on radiation-driven wind models.}
   {The wind efficiency shows the possible presence of a local maximum around an effective temperature of 21~000 K -- in
   qualitative agreement with predictions. Furthermore, we find that the absolute values of the
   radio mass-loss rates show good agreement with empirical H$\alpha$
   rates derived assuming homogeneous winds -- for the spectral range under consideration. However, the empirical mass-loss rates are
   larger (by a factor of a few) than the predicted rates from radiation-driven wind theory for
   objects above the bi-stability jump (BSJ) temperature, whilst they are smaller (by a factor of a few) for stars
   below the BSJ temperature. The reason for these discrepancies remains as yet unresolved. A new wind momenta-luminosity relation (WLR) for O8-B0 stars has been derived using the radio observations. The validity of the WLR as a function of the fitting parameter related to the force multiplier $\alpha_{\rm eff}$ (Kudritzki \& Puls 2000) is discussed.}
   {Our most interesting finding is that the qualitative behaviour of the empirical
    wind efficiencies with effective temperature is in line with the predicted behaviour, and this
    presents the first hint of empirical evidence for the predicted
   $mass$-$loss$ bi-stability jump. However, a larger sample of stars around the BSJ needs to be observed
    to confirm this finding.}
\authorrunning{P. Benaglia et al.}
\titlerunning{The BSJ at radio wavelengths}

   \maketitle
%

\section{Introduction}
Massive stars are the main drivers for the evolution of galaxies: they are the prime
contributors to the energy and momentum input into the interstellar medium through
stellar winds and supernovae, they generate the bulk of ionising radiation, and they are
important sources for the chemical enrichment of carbon, oxygen and nitrogen in the Universe.
Their very evolution through subsequent stages (OB --$>$ LBV --$>$ WR --$>$ SN) is
believed to be {\it driven} by mass loss (e.g. Chiosi \& Maeder 1986).
Heger et al. (2003) have recently reviewed
the way in which massive stars end their lives and they find that the type of remnant
that is left -- neutron star, black hole, or no remnant at all -- is primarily dependent on the
amount of mass lost via stellar winds.

The physical mechanism for these winds of massive OB stars has
long been identified to be that of radiation pressure on
millions of spectral lines (Lucy \& Solomon 1970; Castor Abbott \& Klein 1975: hereafter CAK).
These
radiation-driven wind models predict the mass-loss rates $\dot{M}$
of O stars to depend strongly on the stellar luminosity, and the
terminal flow velocity ($v_\infty$) to be proportional to the
escape velocity ($v_{\rm esc}$), reaching wind velocities of
thousands of km s$^{-1}$. Over the last decades, the theory has
proven to be very successful in explaining the overall mass-loss
properties of O stars, with $\dot{M}$ of O stars up to 10$^{-5}$
M$_{\odot}$\,yr$^{-1}$ and a clear correlation between the wind
terminal and the escape velocity: ($v_\infty / v_{\rm esc}$)
$\simeq$ 2.6 (Kudritzki \& Puls 2000).\\

Nevertheless, there are many open issues in the field of massive
star evolution and their winds -- even during the ``best
understood'' O \& B evolutionary phases. One of them is that of
wind clumping (e.g. Eversberg et al. 1998). Many modern evolution models
(Maeder \& Meynet 2003, Limongi \& Cioffi 2006, Eldridge \& Vink 2006) use OB mass-loss rates from smooth
wind Monte Carlo mass-loss predictions of Vink et al. (2000a,
2000b). However, the sizes of these mass-loss rates have recently
been questioned by a number of ultraviolet (UV) studies, and
downward revisions by factors of five to ten have been suggested
(Bouret et al. 2003, Fullerton et al. 2006). There are theoretical and
observational indications (e.g. Evans et al. 2004) that
the winds of OB stars are clumped, but what remains uncertain is
by {\it what factor} the mass-loss rates could be over-predicted;
is it as dramatic (i.e. factors $\sim$ 10) as has been suggested
by the UV studies, or is the clumping factor relatively small (a
factor of $\sim$ two)?. It is relevant to note that a modest
amount of wind clumping is actually {\it required} to match the
Vink et al. (2000a, 2001) mass-loss predictions. If the clumping
factor ($f$) is only $\sim$ 5 - the corresponding mass-loss rates
are a factor $\sqrt5$ lower- (Repolust et al. 2004, Mokiem et al. 2007), massive
star evolution is not anticipated to be dramatically affected by
clumping. For a recent discussion on wind clumping from optical
studies, we refer to Puls et al. (2006), and Davies et al.
(2007).\\

Another topical issue is that of the ``bi-stability jump'' (BSJ): an
empirical drop in the ratio ($v_\infty / v_{\rm esc}$) from 2.6 to 1.3
around temperatures of about 21~000 K (Lamers et al. 1995). Theoretically, the wind
characteristics as a function of stellar spectral type are best
described in terms of the wind efficiency number: $\eta =
(\dot{M}v_{\infty})/({L_*/c})$, a measure for how much momentum
from the photons is transferred to the ions of the outflowing
wind. Vink et al. (2000a) computed a large grid of wind
models as a function of effective temperature and presented the
following overall behaviour of the mass-loss rate as a function of
decreasing temperature (Fig.~1).

Over the temperature range between 50 000 K down to 27 500 K, the
mass-loss rates drop rapidly. The reason is that of a growing
mismatch between the wavelengths of the maximum opacity in the UV
and the flux maximum, which progressively moves from extreme UV
(Ly continuum) to far UV, from early O to early B stars. The
behaviour is however reversed around the BSJ, where $\eta$ is now
predicted to {\it increase} by a factor of 2-3 to a local maximum
when Fe {\sc iv} recombines to the more efficient Fe~{\sc iii}
configuration (Vink et al. 1999, Vink 2000). Below the BSJ, the first effect
returns, and $\eta$ is again predicted to decrease with
temperature.

   \begin{figure}
   \centering
 \includegraphics[width=8cm]{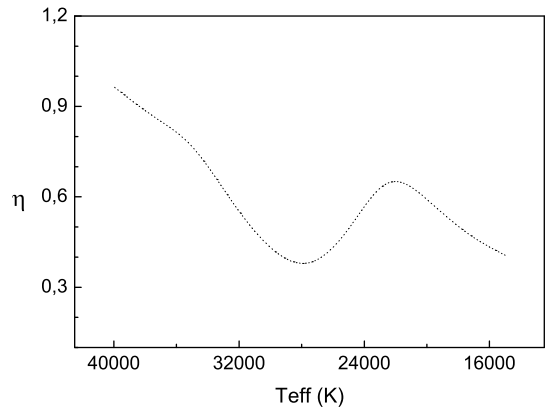}
  \caption{The predicted wind efficiency number: $\eta = (\dot{M}v_{\infty})/({L_*/c})$
as a function of effective temperature (from Vink et al. 2000a). Note
the presence of a local maximum at the position of the
bi-stability jump around $T_{\rm eff} = 21~000$K.}
              \label{etapred-1}%
    \end{figure}

For B supergiants, Vink et al. (2000a) reported significant discrepancies between empirical and predicted 
mass-loss rates.
Recent studies involving H$\alpha$ (Trundle \& Lennon 2005, Crowther et al. 2006) and UV analyses
(Prinja et al. 2005) have reconfirmed these findings.
However, discrepancies between empirical UV rates and radiation-driven wind
models have now also been reported for O stars (see Fullerton et al. 2006).
It is currently unclear whether the reported discrepancies for B supergiants are due
to the assumption of smooth winds (and the neglect of wind clumping) in the
radiation-driven wind model for {\it all} spectral types, or instead that they are somehow related to
the physics of the BSJ.

\subsection{The bi-stability jump}

The bi-stability jump was first discussed by Pauldrach \& Puls
(1990) in the context of their model calculations of an
{\it individual} star, the Luminous Blue Variable (LBV) P Cygni. The first empirical work on the BSJ {\it for a sample} of stars
was that by Lamers et al. (1995), who found a
discontinuity in the ratio ($v_\infty / v_{\rm esc}$) at spectral
type B1 ($T_{\rm eff}$ $\sim$ 21~000 K). For the earlier-type
stars the ratio is $\simeq$ 2.6, but for the lower temperatures it
drops to $\simeq$ 1.3. They also discussed the possible existence
of a second BSJ at spectral type A0. We note that Crowther et al.
(2006) performed a sophisticated analysis using
line-blanketed model atmospheres
(of Hillier \& Miller 1998) of the temperatures of objects around the B1 BSJ
and concluded the empirical BSJ in wind velocity represents a more
gradual decrease than suggested by the step function of Lamers et
al. (1995).

On the theoretical front, Vink et al. (1999) studied the nature of
the winds around the BSJ and found a gradual change in wind
properties, while they attributed the origin of the BSJ to a
gradual change in the ionization of the Fe lines that drive the
wind. For a range of stellar parameters and adopted velocity laws,
Vink et al. (1999, 2000a) found an increase in the wind efficiency
$\eta$ by a factor 2-3 {\it independent} of the assumed terminal
velocity. When accounting for the empirical drop in terminal
velocity, they found a jump in the mass-loss rate by a factor of
five -- for stars at the same luminosity. They argued that the
jump in mass loss is accompanied by a decrease of the ratio
($v_\infty / v_{\rm esc}$), which is the observed bi-stability
jump in terminal velocity. Using self-consistent models they
obtained ratios within 10 percent of the observed values around
the jump (Vink et al. 1999, 2000b; Vink 2000) after a detailed
investigation of the line acceleration for models around the jump,
they demonstrated that $\dot{M}$ increases at the BSJ due to an
increase in the line acceleration of Fe~{\sc iii} near the stellar
photosphere.

The bi-stability mechanism may be the crucial piece of physics for
understanding the least understood phases of massive star
evolution, i.e. the LBV and the B[e] phases. Vink \& de Koter
(2002) found that the bi-stability jump may explain the
mass-loss variability of LBVs. Smith et al. (2004) showed
that, under certain conditions (when stars have lost a lot of mass
due to mass loss), the bi-stability jump may be able to explain
the formation of pseudo-photospheres and possibly even the
existence of Yellow Hypergiants. Kotak \& Vink (2006)
recently suggested that the quasi-sinusoidal modulations in the
radio lightcurves of a number of supernovae may be caused by variable
LBV wind strengths, suggesting that LBVs may be the progenitors of some
supernovae, which would have profound consequences for our
most basic understanding of massive star evolution.

For rapidly rotating stars with radiative envelopes, the pole may
be expected to be hotter than the stellar equator, as due to the
Von Zeipel gravity darkening effect, implying that the BSJ could
provide a higher mass loss from the equator than from the pole for
B stars. This will give rise to a high density ``disk-forming
region'' at the stellar equator -- possibly causing the B[e]
phenomenon (Lamers \& Pauldrach 1991, Pelupessy et al. 2000, Cure et al. 2005).

However attractive the physics of the BSJ may be,
the expected jump in mass loss has yet to be investigated on
observational grounds, which is the prime goal of this study.
We perform a comprehensive radio analysis of massive
stars over the spectral range O8-B3, corresponding to $T_{\rm eff}$ in the critical range
of 35 000 - 15 000 K.\\

The contents of the paper is as follows. Section~2 briefly reviews the
derivation of stellar mass-loss rates by means of radio data.
Section~3 describes the
stellar sample, including the radio observations carried out
specifically for this work. In Sect.~4 we
explain the adopted stellar parameters and the derivation of radio
mass-loss rates. Section~5 presents the wind
efficiencies and a wind momentum-luminosity relation  obtained with radio data.
In Section~6, we compare our
radio results with theoretical predictions.
In Section 7 we analyze the correlation between radio-
and H$\alpha$-derived mass-loss rates.
Section 8 closes with a summary and the conclusions of the present study.

\section{Stellar mass-loss rates from radio observations}

The measurement of stellar mass-loss rates can be achieved by
means of at least three methods, which involve observations at radio,
optical (H$\alpha$), and UV wavelengths. Each technique
retrieves information from a different part of the wind.
Although the radio method is not as sensitive as the H$\alpha$ or UV
methods, the radio flux density method is generally accepted to be the
most accurate (Abbott et al. 1980, Lamers \& Leitherer 1993). The reason is that
the method does not require {\sl a priori} knowledge of the excitation, ionization or
velocity structure in the winds, as the other approaches do to a more
or lesser extent.

Although the radio method is considered to be the most accurate,
it nonetheless depends on the assumption that the radio flux
density is predominantly thermal in nature. This may be quite a
good assumption for the objects under consideration in this study,
as we expect that non-thermal emission at these spectral types
should not be important (even irrespective of the wavelength): all
(13) targets in the range O8-B3 detected at two frequencies by
Bieging et al. (1989), and by Scuderi et al.
(1998) showed thermal spectral indices. Other potential
inaccuracies in the radio mass-loss rates are due to uncertainties
in the stellar distance $d$ and only slightly on the value of the
terminal velocity of the winds. We discuss the effect of wind
clumping in the last Sections.

Despite the weakness of the thermal radio emission from OB stellar
winds (usually less of 1 mJy), the number of closeby ($\leq 3$
kpc) targets is enough to perform a statistical study.
The thermal radio flux density $S_\nu$ at a frequency $\nu$, from an
optically thick stellar wind, is related to the stellar mass-loss
rate $\dot{M}$ as (Wright \& Barlow 1975, Panagia \& Felli 1975):

\begin{equation} 
 \dot{M} = 5.32 \times 10^{-4} \, \frac{(S_{\nu})^{3/4}\,d^{3/2} \,
           v_\infty \, \mu}{Z\, \sqrt{\gamma\, g_\nu \,\nu}} \,\,\,\,\,\,
           {\rm M}_{\odot} {\rm yr}^{-1}~.
\end{equation}

\noindent Here $\mu$ is the mean molecular weight of the ions, $Z$
 the rms ionic charge, $\gamma$ the mean number of electrons per
 ion, and $g_\nu$ the Gaunt factor, approximated by:

\begin{equation} 
 g_\nu = 9.77\, [1 + 0.13\, \log (0.4\,T_{\rm
 eff}^{3/2} (Z \nu)^{-1})  ].
 \end{equation}

\section{Observations at radio wavelengths}

In order to derive mass-loss rates from continuum radio
observations, for stars with effective temperatures around 21~000 K,
we aimed at supergiants of spectral types in the range O8 to B3.
We present here the results of our first observing campaign, with
targets of spectral type spread over the mentioned range. The sample was
completed with the addition of the results from previous
observations taken from the literature and carried out by several authors.\\

The observing campaign consisted of 12 targets with declinations
between $-57^\circ$ to $+37^\circ$. Eight of them were observed
with the NRAO\footnote{The National Radio Astronomy Observatory is
a facility of the National Science Foundation operated under
cooperative agreement by Associated Universities, Inc.} Very Large
Array (VLA), while the southern ones were observed using the
Australia Telescope Compact Array (ATCA)\footnote{The Australia
Telescope Compact Array is funded by the Commonwealth of Australia
for operation as a National Facility by CSIRO.}. The stars can be
identified in Table 1, with the acronyms ATCA or
VLA (third column).
The targets were required to be closeby and have as-accurate-as-possible
determinations of terminal velocities. The integration time was
estimated as the time needed to detect the radio flux density
corresponding to the star mass-loss rate as predicted by Vink et
al. recipe\footnote{http://www.astro.keele.ac.uk/$\sim$jsv/} (Vink 2000).

\subsection{ATCA data}

The ATCA observations were conducted on June 26--27, 2005, at 6B
array configuration. The observing wavelengths were chosen at the
12mm band, i.e. high frequencies, where not only the thermal flux
density is stronger, but also the contribution of non-thermal
emission to the radio flux density may be disregarded. The angular
resolution is about $1''$. The observing frequencies were 17.728
and 17.856 GHz, and the total bandwidth was 128 MHz in each of the
two IFs. Due to bad weather 25\% of the data were not useful, and
had to be flagged. The sources were observed during 6 min scans,
interleaved with 2 min scans of phase calibrator observations. The
bandpass calibrator was 1253-055, and the flux density calibrator
was PKS 1934-638 ($S_{\rm 12mm}$ = 1.03 Jy) during both runs.
Table 1 lists the phase calibrator and the integration time for
each target source.

The data reduction and analysis were performed with the {\sc
miriad} package. One star was detected, HD\,148379, with a flux
density of $0.28\pm0.05$ mJy (ure 2). The synthesized beam
for each source is given in Table 1. Table 2 shows the flux
densities and 1-$\sigma$ errors, and, for the case of
non-detections, the flux density value quoted as an upper limit
corresponds to a 3-$\sigma$ value.

\begin{figure*}[] 
\begin{center}
  \includegraphics[width=120mm]{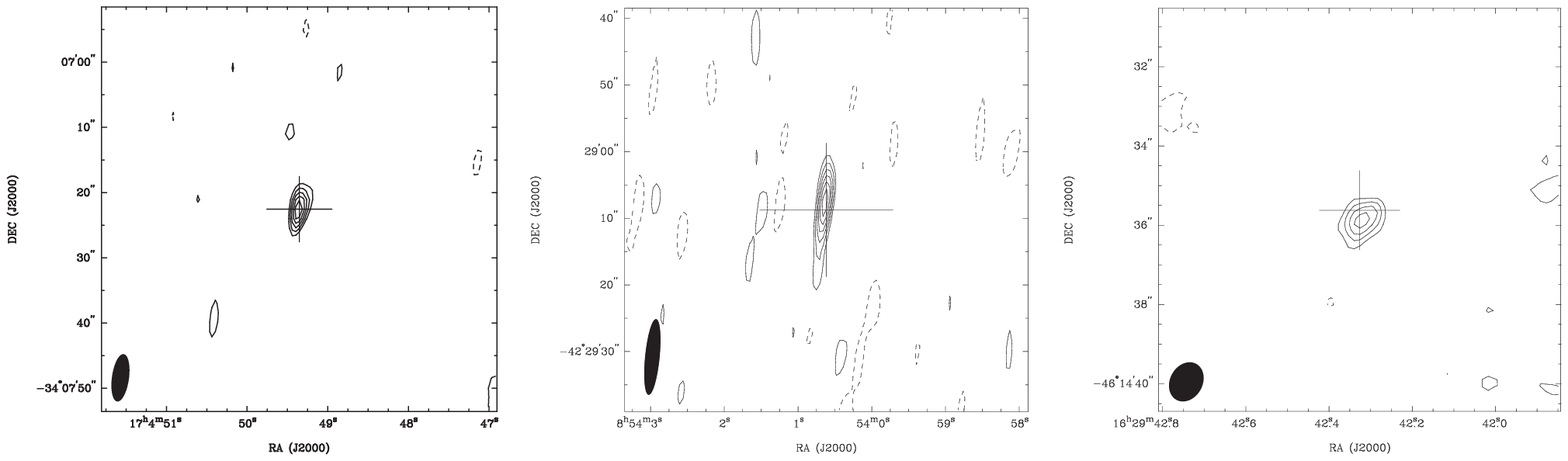}
\caption{{\sl Left panel}: VLA radio continuum image of HD 154090
at 8.46 GHz. Contour levels: --0.08, 0.08 ($2\sigma$), 0.12, 0.15,
0.18. and 0.21 mJy beam$^{-1}$. {\sl Central panel}: VLA radio
continuum image of HD 76341 at 8.46 GHz. Contour levels: --0.15,
0.15 ($3\sigma$), 0.2, 0.25, 0.3 and 0.35 mJy beam$^{-1}$. {\sl
Right panel}: ATCA radio continuum image of HD 148379 at 17.8 GHz.
Contour levels: --0.1, 0.15 ($3\sigma$), 0.2, 0.25, and 0.3 mJy
beam$^{-1}$. The optical position of the star is marked with a
cross, and the synthesized beam is displayed in the bottom left
corner of each image.}\label{RadioImages-2}
\end{center}
\end{figure*}

\begin{table}[htp]
\caption{Observations details} \label{table:1} \centering
\begin{tabular}{l l r r r }     
\hline\hline Star & Phase & Tel. & Int. & Synth.\\
     & Cal.  &      & time &  beam \\
     &       &      & (h)  &   (arcsec)\\
\hline %
HD 42087  & 0559+238  & VLA  & 0.81 & $2.2\times 1.8$ \\ %
HD 43384  & 0559+238  & VLA  & 0.83 & $2.2\times 1.8$ \\%
HD 47432  & 0641-033  & VLA  & 0.75 & $2.5\times 2.3$ \\%
HD 76341  & 0828-375  & VLA  & 0.50 & $11.3\times 2.3$ \\%
HD 112244 &j1326-5256 & ATCA & 3.18 &  $0.8\times 0.4$ \\ 
HD 148379 & 1600-44   & ATCA & 3.58 & $1.0\times 0.8$ \\ 
HD 154090 & 1626-298  & VLA  & 0.75 & $7.3\times 2.6$  \\%
HD 156154 & 1744-312  & VLA  &  0.5 & $7.8\times 2.6$  \\ %
HD 157246 & 1657-56   & ATCA &  6.9 & $0.6\times 0.5$\\ 
HD 165024 &1740-517   & ATCA & 6.36 & $0.7\times 0.5$\\ 
HD 204172 &2052+365   & VLA  & 1.25 & $3.0\times 2.7$  \\ %
BD-11 4586& 1832-105 & VLA  & 0.5 & $3.8\times 2.7$\\ %
\hline
\end{tabular}
\end{table}

\subsection{VLA data}

The VLA targets were observed on September 9 and 17, 2005, at C
array, during a total observing time of 8 h. The flux density calibrator
was 1331+305 ($S_{\rm 3.5cm}=$ 5.23 Jy). The integration time on source, synthesized beam,
and phase calibrator corresponding to each source are given in
Table 1.
The observations were carried out at 3.5 cm with two IF pairs of 50 MHz
bandwidth, to obtain a
reasonable angular resolution with a moderate amount of
calibration time, and lower noise receivers.

The data was reduced and analyzed with the {\sc aips} and {\sc miriad} routines.
Two stars were detected above the noise: HD 76341, and HD 154090,
with signal-to-noise ratio $\geq$ 6 (see Figure 2). The stellar
field of HD 156154 showed weak emission at a 3-$\sigma$ level, and
a 2-$\sigma$ (probable) noise source coincident with the stellar
position. Diffuse emission over the field of HD 42087 precluded
the detection of the stellar wind, also superposed to a 2MASS
source, and the infrared source IRAS 06066+2307.

An extended source was discovered in the stellar field of BD-11
4586, centered at $(RA, Dec)_{\rm J2000} = (18^{\rm h}18^{\rm
m}6.744^{\rm s}, -11^\circ 18' 16.83'')$, with a total flux
density of 2.8 mJy. No correlation at any wavelength was found for
this source in the literature.

\subsection{Previous radio observations}

We searched the literature for continuum radio observations of
late OB supergiants with spectral types
between O8 to B3. Among the contributions considered, the comprehensive
study by Bieging et al. (1989) on OB stars listed
high-resolution radio observations of 16 such supergiants, six of
which were detected. Lamers \& Leitherer (1993) presented
seven radio detections.
Scuderi et al. (1998) obtained radio flux densities -or upper
limits- for ten candidates. We have also included isolated results
from Stevens (2006) (private communication), Benaglia et al.
(2001), and the latest radio flux densities detected by Puls et
al. (2006) on two O9.5 targets.  From the older studies
(Bieging et al. 1989, Lamers \& Leitherer 1993), only detections were included, as the
observations were carried out using the former VLA less sensitive
receivers. The star HD 193237 (observations by Scuderi et al. 1998) is
to be discussed elsewhere amongst LBVs. The star HD 152408 is not
included within the present sample due to its ambiguous
classification as either ``O8 Iafpe'' or ``WN9 ha''
(Bohannan \& Crowther 1999).

The star HD 149404 is listed by Lamers \& Leitherer (1993).
It has been studied by Rauw et al. (2001), and Thaller et
al. (2001), through optical spectroscopy.
They found evidences of a binary system, formed by an O7.5 I(f) and
an ON9.7 I, with a period of $\sim$ 9 d. Most of the radio emission would be produced
in a colliding wind region; the results are consistent with a picture where the
secondary star is undergoing
Roche lobe overflow.
Within this scenario, the radio flux density measured should
come from the region of colliding winds, as well as the winds of both components.
The mass-loss rate derived from the radio flux density is thus only an upper limit and we
disregard from our upcoming analysis this object as well.

\begin{table*}
\begin{minipage}[h]{\columnwidth}
\caption{Supergiants with radio observations: measurements}
\label{table:2}
\centering
{\small
\begin{tabular}{l l l l r l r r r r r }     
\hline\hline
Star &     Sp. Class. & Observ.& $V$, $B$ & $v_{\rm w}$   & $M_{\rm V}$ & $d$  & $e_d$ & $\nu$ &  $S_{\nu}$& $e_S$ \\
     &                &        &        &(km s$^{-1}$)    &    &(kpc) & (kpc) & (GHz) & (mJy)&   (mJy)\\
\hline
HD 42087  &  B2.5 Ib$^a$  &this work  & 5.76, 5.96$^b$&  735$^c$& -6.4$^d$ & 1.4$^{e}$& 0.3 &8.640& $<$  0.14&  \\
HD 43384  &  B3 Iab$^f$   &this work  & 6.26, 6.71$^g$&  760$^c$& -7.2$^d$ & 1.4$^{e}$& 0.3 &8.640& $<$  0.24&  \\
HD 47432  &  O9.7 Ib$^h$  &this work  & 6.22, 6.36$^i$& 1590$^c$& -6.28$^j$& 1.7$^{e}$& 0.4 &8.640& $<$  0.15&  \\
HD 76341  &  O9 Ib$^k$    &this work  & 7.16, 7.46$^i$& 1520$^c$& -6.29$^j$& 1.9$^{e}$& 0.4 &8.640&      0.38& 0.05\\
HD 112244 &O8.5 Iab(f)$^l$&this work  & 5.38, 5.40$^i$& 1575$^c$& -6.29$^j$& 1.5$^{e}$& 0.3 &17.79& $<$  0.3&   \\
HD 148379 &  B1.5 Iape$^m$&this work  & 5.34, 5.80$^n$&  510$^c$& -7.5$^d$ & 1.3$^{e}$& 0.3 &17.79&      0.28& 0.05\\
HD 154090 &  B0.7 Ia$^d$  &this work  & 4.87, 5.13$^d$&  915$^d$& -6.8$^d$ & 1.1$^{d}$& 0.3 &8.640&      0.23& 0.04\\
HD 156154 &  O8 Iab(f)$^l $&this work & 8.05, 8.65$^i$& 1530$^o$& -6.3$^j$ & 2.2$^{e}$& 0.5 &8.640& $<$  0.15&  \\
HD 157246 &  B1 Ib$^m$    &this work  & 3.34, 3.21$^b$&  735$^c$& -6.8$^d$ & 1.1$^{e}$& 0.2 &17.79& $<$  0.18&  \\
HD 165024 &  B2 Ib$^m$    &this work  & 3.66, 3.58$^b$& 1185$^c$& -7.2$^d$ & 0.8$^{e}$& 0.2 &17.79& $<$  0.18&  \\
HD 204172 &  B0 Ib$^h$    &this work  & 5.94, 5.84$^p$& 1685$^c$& -6.4$^d$ & 3.0$^{e}$& 0.7 &8.640& $<$  0.08&  \\
BD-11 4586&  O8 Ib(f)$^q$ &this work  &  9.4, 10.4$^i$& 1530$^o$& -6.3$^j$ & 1.8$^{e}$& 0.4 &8.640& $<$  0.3&   \\
          &              &     &                 &         &          &           &       &     &         & \\
HD 2905   & BC 0.7 Ia$^d$&SPS98&  4.16,  4.30$^d$& 1105$^d$& -7.1$^d$ & 1.1$^{d}$& 0.3 &8.450& 0.4&  0.03\\
HD 30614  &  O9.5 Ia$^d$ &SPS98&  4.29,  4.33$^i$& 1560$^d$& -6.6$^d$ & 1.0$^{d}$& 0.2 &14.95& 0.65& 0.13\\
HD 37128  &  B0 Ia$^d$   &SPS98&  1.70,  1.51$^d$& 1910$^d$& -6.3$^d$ & 0.4$^{d}$& 0.1 &14.95& 1.4&   0.1\\
HD 37742  &  O9.7 Ib$^r$ &LL93 &  1.76,  1.55$^i$& 1860$^c$& -6.3$^j$& 0.4$^{e}$& 0.1 &8.000& 0.89& 0.04\\
HD 41117  &  B2 Ia$^d$   &SPS98&  4.63,  4.91$^d$&  510$^c$& -7.6$^d$ & 1.5$^{d}$& 0.4 &14.95& 0.63& 0.13\\
HD 80077  &  B2 Ia+$^s$  &LCK95&  7.56,  8.90$^t$&  140$^t$& -7.2$^t$ & 3.0$^{p}$& 1.2 &8.640& 0.5&  0.11\\
HD 151804 &  O8 Iaf$^l$  &LL93 &  5.22,  5.26$^u$& 1445$^c$& -6.3$^j$ & 1.9$^{v}$& 0.3 &4.800& 0.4&  0.1\\
HD 152236 &  B1.5 Ia+$^d$&Ste06&  4.73,  5.21$^d$&  390$^d$& -8.8$^d$ & 2.0$^{d}$& 0.3 &8.640& 2.4&  0.1\\
HD 152424 &  OC9.7 Ia$^r$&LL93 &  6.31,  6.71$^i$& 1760$^c$& -6.3$^j$& 1.7$^{e}$& 0.4 &8.000& 0.21& 0.04\\
HD 163181 & BN0.5 Iap$^w$&BCK01&  6.60,  7.15$^x$&  520$^c$& -6.1$^d$ & 1.6$^{e}$& 0.4 &8.640& 0.44& 0.05\\
HD 169454 &  B1 Ia$^y$   &BAC89&  6.65,  7.39$^x$&  850$^z$& -6.8$^d$ & 0.9$^{e}$& 0.2 &15.00& 1.9&  0.1\\
HD 190603 &  B1.5 Ia+$^d$&SPS98&  5.65,  6.19$^d$&  485$^d$& -7.5$^d$ & 1.5$^{e}$& 0.4 &14.95& $<$ 0.7&  \\
HD 194279 &  B2 Ia$^d$   &SPS98&  7.05,  8.07$^d$&  550$^d$& -7.0$^d$ & 1.2$^{d}$& 0.3 &8.450& 0.44& 0.1\\
HD 195592 &  O9.5 I$^l$  &SPS98&  7.08,  7.95$^i$& 1765$^o$& -6.3$^j$& 1.0$^{e}$& 0.2 &14.95& 0.9&  0.13\\
HD 198478 &  B2.5 Ia$^d$ &SPS98&  4.86,  5.28$^d$&  470$^d$& -6.4$^d$ & 0.8$^{d}$& 0.2 &8.450& $<$  0.27&    \\
HD 209975 &  O9.5 Ib$^z$ &PMS06&  5.10,  5.18$^i$& 2050$^z$& -5.45$^z$& 0.8$^{z}$& 0.2 &14.94& 0.422& 0.12\\
CygOB2-10 &  O9.5 I$^z$  &PMS06&  9.88  11.47$^\alpha$& 1650$^z$& -6.95$^z$& 1.7$^{z}$& 0.2 &14.94& 0.3&   0.1\\
BD-14 5037&B1.5 Ia$^\beta$&SPS98&8.22,  9.58$^x$&  750$^o$& -7.4$^o$ & 1.0$^{e}$& 0.2 &14.95& 1.1&   0.2\\
\hline
\end{tabular}
}
\end{minipage}
\medskip\\
$^a$Perryman et al. 1997, $^b$Johnson et al. 1966,
$^c$Howarth et al. 1997, $^d$Crowther et al. 2006,
$^e$derived with {\sc chorizos}: see text,
$^f$Lesh et al. 1968,
$^g$Moffett \& Barnes 1979,
$^h$Walborn 1976, $^i$GOS Catalogue (Ma\'{\i}z Apell\'aniz et al. 2004),
$^j$Martins et al. 2005, $^k$Garrison et al. 1977,
$^l$Walborn 1973,
$^m$Hiltner et al. 1969,
$^n$Feinstein \& Marraco 1979,
$^o$Prinja et al. 1990,
$^p$Crawford et al. 1971,
$^q$Walborn 1982,
$^r$Walborn 1972a,
$^s$Leitherer et al. 1995 (LCK95),
$^t$Schild et al. 1983,
$^u$Bohannan \& Crowther 1999,
$^v$Humphreys 1978,
$^w$Walborn 1972b,
$^x$Kozok et al. 1985,
$^y$Bieging et al. 1989 (BAC89),
$^z$Puls et al. 2006 (PMS06),
$^\alpha$Massey \& Johnson 1991,
$^\beta$Morgan et al. 1955.
SPS98: Scuderi et al. 1998.
LL93: Lamers \& Leitherer 1993.
Ste06: I.R. Stevens (private communication).
BCK01: Benaglia et al. 2001.\\
\end{table*}

\subsection{Considerations on radio-detectability}

The O8 to B3 supergiant sample comprises 19 detections (up to 3
mJy) plus upper limits to 11 sources (typically below 0.3 mJy).
With the aim of comparing the different radio flux densities
measured towards the sample stars, we have extrapolated the values
of $S_\nu$ to a unique frequency of 14.95 GHz, assuming that the
thermal flux behaves as $S_\nu \propto \nu^{0.7}$ (Setia Gunawan
et al. 2000, Williams 1996). In Figure 3 we have plotted the
quantity $S_\nu*(d/{\rm 1kpc})^2 \propto L$ as a function of the
stellar effective temperature. We found no correlation of the
corrected flux with stellar effective temperature, although the
errors in stellar distances are strongly affecting the result.

No correlations between radio flux density and terminal velocity, distance or visual magnitude were found, as expected.

   \begin{figure}
   \centering
  \includegraphics[width=8cm]{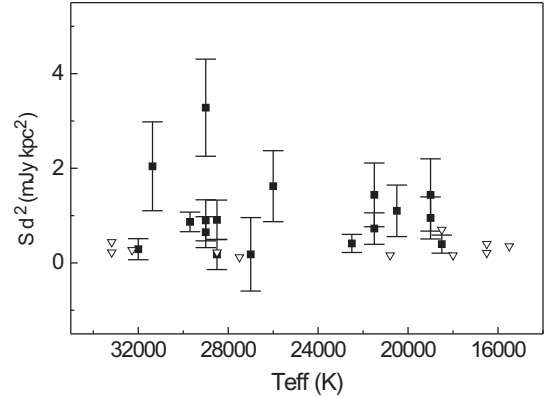}
  \caption{Radio flux density corrected for the stellar distance, as a function of effective temperature.
   The squares represent detected stars, whilst the triangles
represent radio flux density upper limits. The star HD 80077 was
excluded due to its large error in distance ($\sim$40\%).}
              \label{Detect-3}%
    \end{figure}

\section{Stellar parameters and mass-loss rates}

Table 2 lists the 30 selected stars in two groups (observed by us,
and compiled from the literature), ordered by HD number, 
together
with other observational stellar parameters.

The spectral types for O stars were taken from the GOS catalogue
 (Ma\'{\i}z Apell\'aniz et al. 2004), Crowther et al. 
(2006), and Puls et al. (2006). The various
references on spectral types of B stars can be seen in Table
2. The $V,B$ and absolute adopted magnitudes are also given.

Most of the stellar wind terminal velocities quoted were published by
Howarth et al. (1997). 
For four objects, the values were
taken from the interpolations by Prinja et al. (1990), their Table 3, and involved relatively
large errors. For the rest, we have adopted errors of 10\%.\\

   \begin{figure}
   \centering
 \includegraphics[width=8cm]{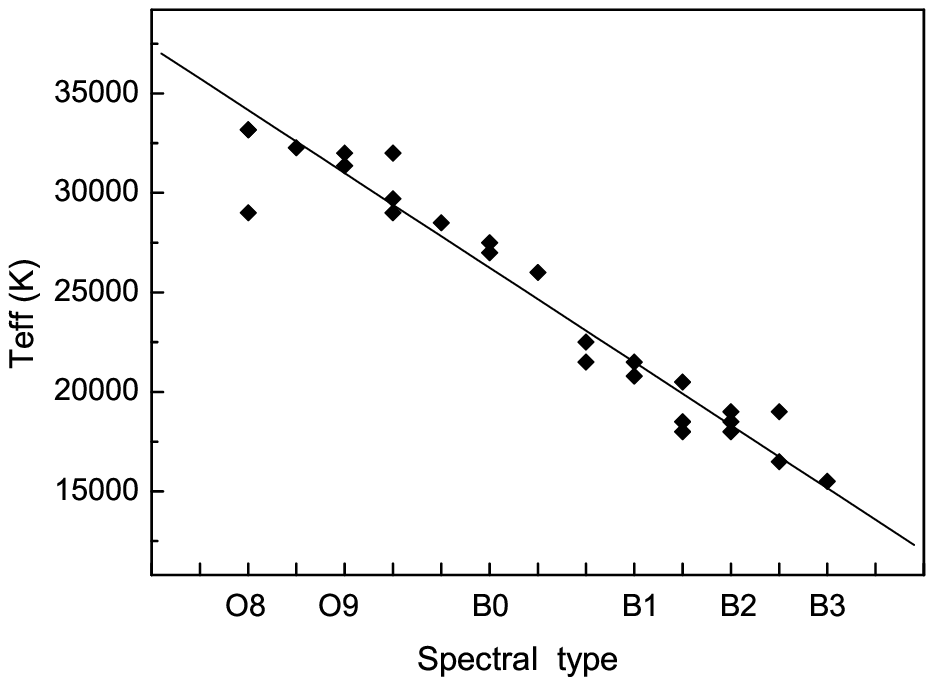}
  \caption{The adopted stellar effective temperatures as a function of spectral type, for OB supergiants.
  The straight line represents the fit $y = 37327(\pm 387) - 1583(\pm 44)x$, where $x$ is an index running on
the different spectral types (from 2 to 14), and $y = T_{\rm eff}$.}
              \label{Teffscale-4}%
    \end{figure}

To achieve the goal of a uniform database,
we have established a common temperature scale for O8-B3 supergiants, re-derived the stellar
distances, and calculated the mass-loss rates for all stars in the
sample using those consistent parameters.\\

The effective temperature scale was established taking into
account the latest work by Crowther et al. (2006) for individual
cases whenever possible, and also from interpolation using their
Table 4. For some O supergiants the values were taken from the
models by Martins et al. (2005). Isolated values are from Lamers
\& Leitherer (1993) and Puls et al. (2006). The resulting scale is
shown in Figure 4. We have assigned to the thirteen different
spectral types and subtypes covered, from O8 to B3, a correlated
number (from 2 to 14). Figure 4 plots the stellar effective
temperature as a function of this integer number. Spectral types
were used to derive temperatures for the stars in the sample. The
uncertainties introduced by the discrete nature of the spectral
classification scheme, the current state of the art of hot-star
atmosphere modeling, and the possible slight mis-classifications
are expected to be of the order of 2000 K for late-O stars and
1500 K for early-B stars. The effective temperature adopted for
each star is listed in Table 3.

References for the adopted stellar luminosities are quoted in Table 3.  
The effective temperature and luminosity of HD 42987, HD 148379, and HD 151804 have been
derived specifically for this work by means of H$\alpha$ profile fitting (see Section 7).

In order to derive
the stellar masses, needed in turn to obtain predicted mass-loss rates (see Section 6), we used the
calibration of surface gravity ($g$) with effective temperature published by Crowther et al. (2006)
 (see their Figure 1, which was built using SMC data presented by Trundle \& Lennon 2004).
The likely range in the stellar mass for each star can be derived allowing a deviation by up to 0.2 dex in $\log(g)$.

\begin{table*}
\caption{Adopted stellar parameters and results} \label{table:3}
\centering
\begin{tabular}{l l l r l r r r r r r } 
\hline\hline
Star & Sp. type & $T_{\rm eff}$ & $\log(L_*/{\rm L}_\odot)$& $log(g)$ & $M_*$ &  $\eta$ & $\dot{M}$ & $e_{\dot{M}}$  & $\dot{M}_{\rm pred}$ & $\dot{M}_{\rm H\alpha}$ \\
     &          &   (K)         &                          &          &  (M$_\odot$)   &&
      \multicolumn{4}{c}{(\,\,\,$10^{-6}\,\,\,\,\,\,$M$_\odot\,\,\,\,\,\,$ yr$^{-1}\,\,\,$)} \\
\hline
HD 156154 & O8 Iab(f)&  33179$^a$ &  5.68$^a$& 3.10& 19.2 &            & $<$ 4.4&      &    3.2&       \\
HD 151804 &  O8 Iaf  &  29000$^b$ &  5.88$^b$& 3.00& 41.4 &$1.0\pm0.12$&    10.4&  3.70&    3.2&  16.0$^b$ \\
BD-11 4586&  O8 Ib(f)&  33179$^a$ &  5.68$^a$& 3.10& 19.2 &            & $<$ 5.6&      &    3.2&   (2.6)$^c$ \\
HD 112244 &O8.5 Iab(f)& 32274$^a$ &  5.65$^a$& 3.10& 20.0 &            & $<$ 3.0&      &    2.6&       \\
HD 76341  &  O9 Ib   &  31368$^a$ &  5.61$^a$& 3.05& 18.2 &$1.3\pm0.17$&     7.2&  2.51&    2.1&       \\
CygOB2-10 &  O9.5 I  &  29700$^d$ &  5.82$^d$& 3.00& 32.8 &$0.5\pm0.07$&     4.4&  1.77&    2.7&   2.7$^{d,e}$ \\
HD 195592 &  O9.5 I  &  29000$^f$ &  5.60$^f$& 3.00& 21.7 &$1.0\pm0.14$&     4.8&  1.69&    1.2&   3.8$^d$ \\
HD 30614  &  O9.5 Ia &  29000$^f$ &  5.63$^f$& 3.00& 23.3 &$0.6\pm0.08$&     3.3&  1.22&    1.5&   5.0$^f$ \\
HD 209975 &  O9.5 Ib &  32000$^d$ &  5.31$^d$& 3.10&  9.5 &$1.2\pm0.15$&     2.4&  0.94&    0.7&   1.1$^d$ \\
HD 37742  &  O9.7 Ib &  28500$^f$ &  5.57$^a$& 2.95& 19.4 &$0.4\pm0.05$&     1.5&  0.49&    0.9&   (1.8)$^g$ \\
HD 47432  &  O9.7 Ib &  28500$^f$ &  5.57$^a$& 2.96& 19.8 &            & $<$ 3.1&      &    1.1&   (1.9)$^h$ \\
HD 152424 &  OC9.7 Ia&  28500$^f$ &  5.57$^a$& 2.96& 19.8 &$1.0\pm0.14$&     4.5&  1.65&    1.0&       \\
HD 37128  &  B0 Ia   &  27000$^f$ &  5.44$^f$& 2.90& 15.9 &$0.5\pm0.07$&     1.6&  0.54&    0.5&   2.3$^f$ \\
HD 204172 &  B0 Ib   &  27500$^f$ &  5.50$^f$& 2.93& 18.2 &            & $<$ 5.0&      &    0.7&       \\
HD 163181 &  BN0.5 Iap& 26000$^f$ &  5.57$^a$& 2.85& 22.4 &$0.1\pm0.02$&     2.1&  0.73&    2.5&       \\
HD 154090 &  B0.7 Ia &  22500$^f$ &  5.48$^f$& 2.60& 18.1 &$0.2\pm0.03$&     1.3&  0.48&    3.2&   1.0$^f$ \\
HD 2905   &  BC0.7 Ia&  21500$^f$ &  5.52$^f$& 2.53& 20.3 &$0.4\pm0.05$&     2.5&  0.84&    2.3&   2.0$^f$ \\
HD 169454 &  B1 Ia   &  21500$^f$ &  5.45$^f$& 2.53& 17.3 &$0.5\pm0.06$&     3.4&  1.14&    2.9&       \\
HD 157246 &  B1 Ib   &  20800$^i$ &  5.45$^f$& 2.45& 16.4 &            & $<$ 0.6&      &    3.4&       \\
BD-14 5037&  B1.5 Ia &  20500$^f$ &  5.60$^f$& 2.45& 24.5 &$0.2\pm0.04$&     2.5&  1.18&    4.9&   (3.4)$^j$ \\
HD 152236 &  B1.5 Ia+&  18000$^f$ &  6.10$^f$& 2.20& 73.4 &$0.1\pm0.02$&     8.3&  2.70&   41.7&     6$^f$ \\
HD 190603 &  B1.5 Ia+&  18500$^f$ &  5.57$^f$& 2.25& 21.8 &            & $<$ 2.1&      &    6.8&   2.5$^f$ \\
HD 148379 &  B1.5 Iape& 18500$^b$ &  5.56$^b$& 2.25& 21.3 &$0.1\pm0.01$&     0.8&  0.29&    5.8&   1.5$^b$ \\
HD 194279 &  B2 Ia   &  19000$^f$ &  5.37$^f$& 2.30& 13.9 &$0.2\pm0.02$&     1.5&  0.58&    3.3&   1.1$^f$ \\
HD 41117  &  B2 Ia   &  19000$^f$ &  5.65$^f$& 2.30& 26.4 &$0.1\pm0.02$&     2.1&  0.76&    8.5&   1.0$^f$ \\
HD 80077  &  B2 Ia+  &  18500$^f$ &  5.40$^f$& 2.25& 14.7 &$0.1\pm0.01$&     1.7&  0.69&   30.2&       \\
HD 165024 &  B2 Ib   &  18000$^f$ &  5.40$^f$& 2.20& 14.7 &            & $<$ 0.6&      &    0.9&       \\
HD 198478 &  B2.5 Ia &  16500$^f$ &  5.03$^f$& 2.05&  6.3 &            & $<$ 0.5&      &    1.1&   0.2$^f$ \\
HD 42087  & B2.5 Ib  &  19000$^b$ &  5.08$^b$& 2.30&  7.1 &            & $<$ 1.1&      &    0.9&   0.2$^b$ \\
HD 43384  & B3 Iab   &  15500$^f$ &  4.88$^h$& 1.95&  4.4 &            & $<$ 1.7&      &    0.3&   (0.3)$^h$ \\
 \hline
\end{tabular}
\medskip\\
$^a$Martins et al. 2005, $^b$this work, $^c$Scuderi et al. 1992,
$^d$Puls et al. 2006, $^e$Herrero et al. 2002, $^f$Crowther et al.
2006, $^g$Lamers \& Leitherer 1993, $^h$Morel et al. 2004,
$^i$Lamers et al. 1995, $^j$Scuderi et al. 1998.\\
\end{table*}

Distances were derived using {\sc chorizos}  (Ma\'{\i}z Apell\'aniz 2004) by
comparing the observed with $BVJHK_s$ photometry with low-gravity Kurucz
atmospheres. The temperature for each star was taken from Table 3 and both
$E_{4405-5495}$ and $R_{5495}$ (the monochromatic equivalents to $E_{B-V}$ and
$R_V$, respectively, were left as free parameters. {\sc chorizos} is a Bayesian
code that finds all the solutions for the general inverse photometric
problem in which $N$ parameters (two in this case) are fitted to $M$
independent colors (three in this case). After applying {\sc chorizos} to obtain
the unextinguished photometry, the spectroscopic distance was calculated
using the values of $M_V$ in Table 2, assuming uncertainties between 0.5 and
1.0 magnitudes depending on the spectral type. In all cases we found that
the largest contribution to the final uncertainty in the distance, $e_d$, is
that of the uncertainty in $M_V$ and not that of the photometry itself.

We investigated the membership of the sample stars to stellar clusters or OB associations.
For those stars belonging to stellar groups, we found that in most cases,
our obtained individual distances agree well between errors, with those of the clusters/associations.
For the star HD 151804, catalogued as a member of Sco OB1 
(Humphreys 1978, $d = 1.91$ kpc),
the association distance was adopted.
Very recently, the distance to HD~169454 was established as 1.36 kpc (Hunter et al. 2006). For the sake of
consistency, we have adopted the distance derived here.\\

In order to derive the radio mass-loss rate values for previously
observed stars, we have taken into account the radio flux density
at the highest frequency whenever possible (see Table 2).

All mass-loss rates values were derived using Eq. (1), and
assuming the measured flux density is thermal. This looks like a
reasonable hypothesis, as all but one star (HD 151804) were
observed at $\nu \geq 8$ GHz. Besides, the sample stars that have
been observed by Bieging et al. (1989) and Scuderi et al. (1998)
at more than one frequency, have shown thermal emission. For all
but one star we have adopted abundances of $Y_{\rm H} = 0.85$,
$Y_{\rm He} = 0.15$, and assumed that at the temperatures
considered, all H is ionized, while He is singly ionized.
Consequently $\mu = 1.45$, $Z = \gamma = 1$. The differences in
the mass-loss rate results in considering $\mu =
1.3$ -- 1.6 were lower than 12\%. If the He is doubly ionized 
in O stars, the mass-loss rates derived here are underestimated
by about 30\%, although we find this unlikely given the stellar
effective temperatures involved. For HD 151804, we adopted $Y_{\rm
He} = 0.25$. Errors in $\mu$, $Z$, and $\gamma$ were taken as
10\%; $\sigma_L = 0.5L_*$. Mass-loss rate values and the
corresponding errors derived from error propagation in Eq.~(1) are
listed in Table 3, and plotted in Fig. 5. In the case of the non-detections, a 3-$\sigma$
flux density upper limit was used, and the corresponding mass-loss
rates are upper limits.

   \begin{figure}
   \centering
  \includegraphics[width=8cm]{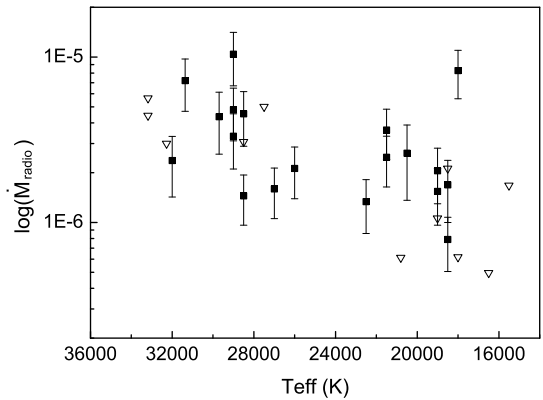}
  \caption{
Mass loss rates derived from radio continuum observations, as a function of effective temperature (squares), and upper limits (open triangles).
}
              \label{mdot-t}%
    \end{figure}

\section{Wind efficiencies}

In order to better compare radio results with results from other
wavelengths or models, we proceeded to compute the wind
efficiencies corresponding to the radio mass-loss rates. The
values are shown in Table 3, and in Figure 6 with filled squares. 
At $T_{\rm
eff} \approx 21.5$ kK there are two (squares) stars with wind efficiencies
well above those of their neighbours: HD 169454 and HD 2905.
The star HD 169454 has been represented twice, for wind efficiencies derived according 
to two possible distances 
(see Sect. 4): with a square when the distance adopted here is used, and with a 
triangle when the Hunter et al. (2006) value is used.

   \begin{figure}
   \centering
 \includegraphics[width=8cm]{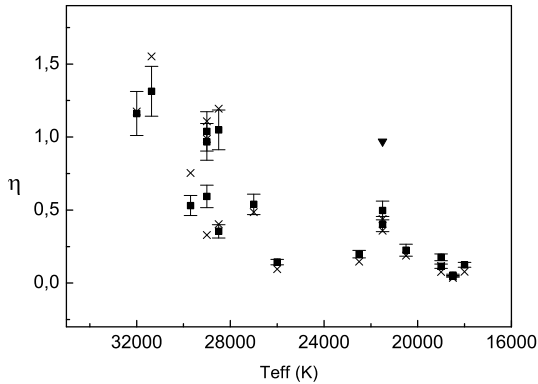}
  \caption{
  Observed wind efficiency versus effective temperature (squares).
  Crosses represent `corrected' values (see text).
The triangle corresponds to the star HD 169454 if the Hunter et al.
(2006) distance is considered. Note the possible presence of a
local maximum at the predicted position of the BSJ.
}
              \label{etaTradio-5}%
    \end{figure}

We wish to check if the empirical shapes of the wind efficiency could be
an artifact of the fact that we observe objects with different
masses and luminosities.
This necessarily introduces a scatter in $\eta$, but we may be able to
correct for this.
The correction was derived in the following manner.
From models for different parameter sets of ($M_*$, $L_*$), we obtain the corresponding
wind efficiencies. 
We choose a parameter set for which $\eta=1$ for the maximum
effective temperature and $v_\infty / v_{\rm esc}$ ratio
(see Fig.~6 from Vink et al. 2000a). The $\eta$ values were plotted
as a function of their Eddington factor ($\propto M_*/L_*$),
resulting in Eddington-corrected $\eta$ values (Fig.~7). For each
target star we can compute the $\eta$ correction value according
to its Eddington factor, and divide the uncorrected $\eta$ by the
correction value. The results are also shown in Figure 6. Though
an additional source of scatter is being introduced as a result of
this via the errors in stellar masses (not shown in the Figure),
the `corrected' wind efficiencies are similar to the original
values. Consequently, hereafter we will use non-corrected values.

   \begin{figure}
   \centering
 \includegraphics[width=8cm]{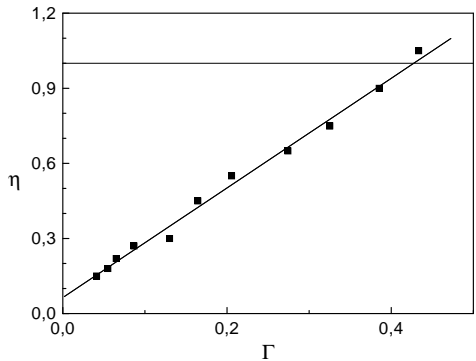}
  \caption{Wind efficiencies taken for the twelve models from Fig.~5 from Vink et al. (2000a),
  as a function of the Eddington factor $\Gamma$. The straight line represents unity.}
              \label{eddincorr-6}%
    \end{figure}

\subsection{The empirical wind-momentum relation}

The radio observations allowed us to estimate the mass-loss rates
for the 19 detected stars in a uniform way. We have utilised this
database to provide a radio derived calibration for the
coefficients $D_{0},x$ of the wind momenta-luminosity relation
(WLR; see Kudritzki \& Puls 2000) for late O (O8-O9) and B0
supergiants. Our WLR is shown in Figure 8. If the wind momenta
equal $D=\dot{M}_* \, v_\infty \, (R/{\rm R}_\odot)^{1/2}$, and
$\log(D) = D_{0} + x \log(L/{\rm L}_\odot)$, a linear fit yields
the values $D_{0}=26.6\pm2.1$; $x=1.0\pm0.4$ when only the O stars
(O8-O9) are included, and $D_{0}=25.0\pm2.7$; $x=1.3\pm0.5$ when
the B0 stars are also included in the fit. For obvious reasons B1
stars are excluded as they are located at the BSJ, whilst for
spectral types later than B1, we have too few datapoints to derive
a reliable WLR.

We note the shallow slope of our derived WLR. For early-type O stars the slope, $x$, is significantly larger.
In standard radiation-driven wind theory, the slope, $x$, is the reciprocal value of the ``effective'' CAK
force multiplier $\alpha_{\rm eff}$ (see Kudritzki \& Puls for definitions). For early O-stars, this value
is approximately 2/3, the reason for which the stellar mass cancels and the WLR can be used in the first place.

Here, we find an empirical $\alpha_{\rm eff}$ in the range 0.78-1 for O8-B0 stars, which may be consistent with the
high $\alpha$ value of $\alpha^{\rm MC} = 0.85$ predicted through Monte Carlo simulations by Vink et al. (1999) for models
at $T_{\rm eff} = 30~000$K. A high value of $\alpha_{\rm eff}$
may suggest a relatively large value for the ratio of ($v_{\infty}/v_{\rm esc}$). Interestingly objects
in this temperature range indeed show the largest ($v_{\infty}/v_{\rm esc}$) ratio according to Fig.~6 of
Kudritzki \& Puls (2000). We note however that for these large values of $\alpha_{\rm eff}$ (when $\alpha_{\rm eff}$
no longer equals 2/3), the concept of the WLR may lose its meaning. Investigations of the
wind efficiency $\eta$ versus temperature may therefore be a more valuable tool to
understand wind driving for different effective temperature regimes.

   \begin{figure}
   \centering
 \includegraphics[width=8cm]{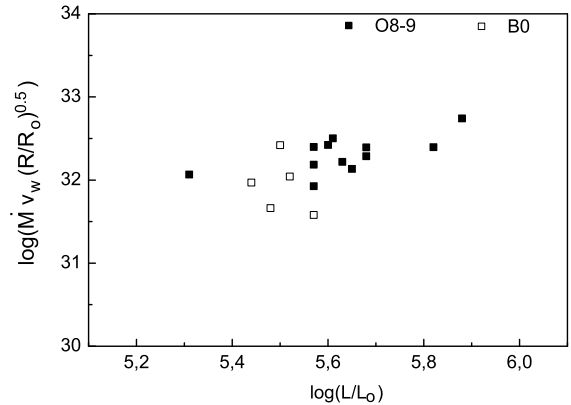}
  \caption{Wind momenta $D$ of galactic late O-, and early B supergiants as a function of luminosity, derived using radio
  continuum observations.
}
              \label{wlr-fit}%
    \end{figure}

\section{Mass-loss rates: expected versus predicted}

Once we have estimated the wind efficiencies from mass-loss rates derived
using radio flux densities, we proceeded to compare the results with model predictions.
Vink (2000), and Vink et al. (2000a) developed mass-loss rate
models and presented them for a $(L_*,M_*)$ grid. Based on the
models, they produced a recipe to derive a stellar mass-loss rate
if the stellar mass, luminosity, effective temperature and
terminal velocity are known. We have made use of the routine to
derive the expected mass-loss rates for the radio
observed stars. The program also provides information on the position of the star
with respect to the BSJ, according to the specific stellar parameters.
The predicted mass-loss rates $\dot{M}_{\rm pred}$ are given in Table 3.

The recipe to derive the predicted mass-loss rates has helped to
discriminate which stars are above and below the BSJ temperature, and we
divided them into two groups. Both groups consist of 15 stars. The
first group contains: HD 30614, HD 37128, HD 37742,
 HD 47432,  HD 76341, HD 112244, HD 152424,
HD 156154, HD 151804, HD 163181, HD 195592,  HD 209975, HD 204172,
BD-11 4586, and  Cyg OB2 No. 10.
The second group contains the remaining 15 objects.

The errors involved in the derivation of the predicted mass-loss
rates are $\sigma_{L_*} = 0.5 \,L_*$, 0.2 dex in $\log(g)$, and
$\sigma_{T {\rm eff}} = 2000 - 1500$ K. A direct comparison
between expected and measured values is presented in Figure 9. We
show the wind efficiencies derived using radio observations
towards Galactic stars, and, superposed, the predicted wind
efficiency number constructed for $\log (L_*/{\rm L_\odot}) =
6.0$, $M_*= 60$ M$_\odot$, $v_\infty / v_{\rm esc} = 2.6$, as in Fig. 1.

   \begin{figure}
   \centering
 \includegraphics[width=8cm]{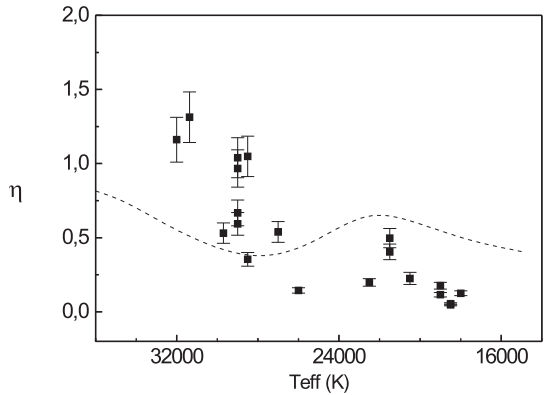}
  \caption{Modelled wind efficiency for $\log (L/{\rm L_\odot}) = 6.0$, $M_*= 60$ M$_\odot$,
$v_\infty / v_{\rm esc} = 2.6$ (dashed line, same as in Fig. 1) superposed to the
wind efficiencies derived by means of radio observations, for
Galactic stars (squares).}
              \label{both-etas}%
    \end{figure}

Our results show that the stars above the BSJ temperature have predicted mass-loss rates lower than the empirical
radio mass-loss rates. The reverse holds true for the stars below the BSJ temperature.
If the radio mass-loss rates are affected by significant wind clumping, these empirical
rates may need to be reduced by a certain amount, and possibly be brought into agreement
with the predictions of radiation-driven wind models.
%
If for stars above the BSJ temperature, the 
discrepancy between the empirical rates and the theoretical rates is attributable to wind clumping 
overestimating empirical rates -- whilst the theoretical rates are not 
affected -- then, adopting a similar clumping correction for stars below 
the BSJ temperature, would worsen the discrepancy between the empirical and 
theoretical rates for B supergiants.\\

The reader should be aware that our analysis has been performed using {\it derived} mass-loss
rates only. ``Upper limits'' are not taken into account in the comparison between empirical radio
rates and mass-loss predictions. Therefore our comparison on the absolute value of the mass-loss rate is necessarily incomplete.
Having noted this, there is no {\it a priori} reason why incompleteness should be a function of
stellar temperature, and although the issue of the absolute mass-loss rates of hot-star winds is still an open one,
the qualitative comparison between the empirical radio mass-loss rates and the Monte Carlo predictions
of radiation-driven wind models appears to support the predicted {\it mass-loss} BSJ. More data around the BSJ
will be necessary to confirm this.

\section{Comparison of radio and H$\alpha$ results}

Many of the stars composing our sample have mass-loss rates in
common with measurements from H$\alpha$ profiles. The values of
$\dot{M}_{\rm H \alpha}$ are listed in Table 3, with their
references (Lamers \& Leitherer 1993; Scuderi et al. 1992, 1998;
Puls et al. 2006; Crowther et al. 2006; Morel et al. 2004; Herrero
et al. 2002).

Additionally, three new H$\alpha$ mass-loss rates have been
derived specifically for this work. The stars HD 42087, HD 148379,
and HD 151804 were analyzed in an identical manner to Crowther et
al. (2006). For the first two stars the analysis was based on
VLT/UVES data (PoP Survey, Bagnulo et al. 2003). For HD 151804,
AAT/UCLES spectroscopy was used, and the data was re-analyzed
using current version of code plus TLUSTY velocity structure at
depth. Figure 10 displays the H$\alpha$ profile of HD 148379 with
errors, showing the reliability with which mass-loss rates can be
achieved.

   \begin{figure}
   \centering
 \includegraphics[angle=-90,width=8cm]{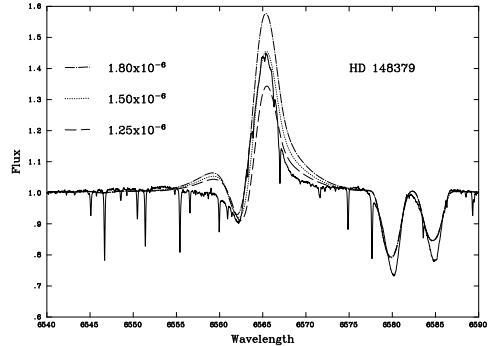}
  \caption{H$\alpha$ profile of HD 148379. Source: UVES PoP Survey.}
              \label{ha-profile}%
    \end{figure}

Figure 11 presents the wind efficiencies derived using mass-loss
rates from H$\alpha$ profiles for 20 stars out of the present
sample (see Table 3). Results from for modern {\sc cmfgen} and
{\sc fastwind} profile fitting are represented with filled
squares. The rest (open squares) are specified in Table 3 with
brackets. The results with $T_{\rm eff}$ are more scattered than
they were for the radio wind efficiencies, but they do not
contradict the predicted behaviour.

   \begin{figure}
   \centering
 \includegraphics[width=8cm]{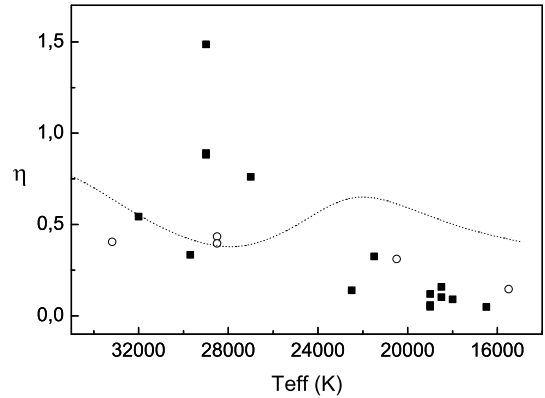}
  \caption{
 Wind efficiencies from H$\alpha$ measurements for the sample stars.
 Filled squares: results from {\sc cmfgen}/{\sc fastwind} profile fitting.
 Open squares: results from other methods. Dashed line: same as Figure 1.}
              \label{mdothalpha}%
    \end{figure}

An analogous plot to the Galactic H$\alpha$ one can be constructed
for extragalactic stars. Trundle et al. (2004), and Trundle \& Lennon
(2005) have obtained the mass-loss rates of 16 B
supergiants earlier than B4 in the Small Magellanic Cloud, by means of
H$\alpha$ measurements, all treated in a homogeneous way.
These $\dot{M}_{{\rm H}_\alpha}$ are
shown in Figure~12, as a function of the temperature scale adopted in this work.
Possibly, the wind efficiencies for some stars around 22~000 K are above the overall decreasing trend, but
larger samples are required to test if the SMC mass-loss rates indeed have a local maximum around the BSJ.

   \begin{figure}
   \centering
 \includegraphics[width=9cm]{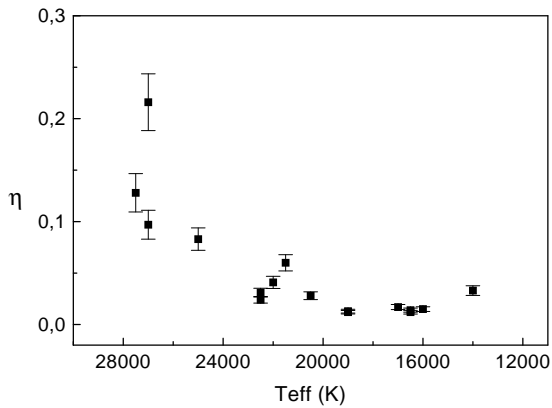}
  \caption{
 Wind efficiencies from H$\alpha$ measurements for SMC stars
  (from Trundle et al. 2004, and Trundle \& Lennon 2005).
}
              \label{mdotha-smc}%
    \end{figure}

Studies of the comparison between mass-loss rates derived using
different methods have been conducted by various groups. Very
recently, Fullerton et al. (2006) derived mass-loss rates from UV
profiles and investigated whether the UV rates matched those
estimated from H$\alpha$ profiles and radio observations. They
found that the results follow the relation:
$\dot{M}_{\rm UV} < \dot{M}_{\rm H\alpha} < \dot{M}_{\rm radio}$
and attributed this discrepancy to the presence of wind
clumping.\\

In the present study we have assumed that the free-free radio
emission is due to smooth winds. If the winds are significantly
clumped, the radio-rates derived in our paper are likely
overestimates (see e.g. Blomme et al. 2003). We constructed
Figure~13 with all radio and H$\alpha$ mass-loss rate results
available. From the figure it is clear that there is a good match
between both sets of data, although the H$\alpha$ lines are formed
in the inner wind, whilst the radio emission arises from the outer
wind. This agreement does not necessarily discard clumping, but it
may support the hypothesis that the clumping structure is
uniformly distributed throughout the whole extension of the wind.

A similar agreement between the inner and the outer wind was found
for a large sample of O stars by Lamers \&  Leitherer (1993).
However, these results have been challenged by Fullerton et al.
(2006) and Puls et al. (2006).

   \begin{figure}
   \centering
 \includegraphics[width=9cm]{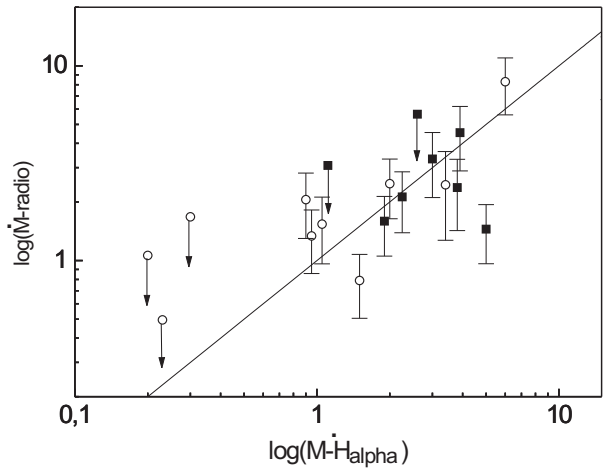}
  \caption{H$\alpha$-derived versus radio-derived mass-loss rates. Units are 10$^{-6}$ M$\odot$ yr$^{-1}$.
  Squares: stars above the
  BSJ temperature; open circles: stars below the jump temperature. Symbols with arrows represent upper limits in
  $\dot{M}_{\rm radio}$.}
              \label{mradmha}%
    \end{figure}

\section{Summary and conclusions}

Throughout the present investigation, a database of radio data of 30 O8-B3 supergiants has been
constructed, with observations carried out by different authors. The stellar distances have been reviewed
and re-derived, using a subroutine that takes into account the IR and optical indices. We have defined a
uniform stellar temperature scale for the spectral range O8-B3.

We have performed high resolution continuum radio observations
towards 12 O8-B3 supergiants, detecting three of them: HD 76341,
HD 148379 and HD 154090, and derived their mass-loss rates. The
rates of another 18 stars were re-determined using results from
the literature.

New H$\alpha$ mass-loss rates have been obtained for three objects: HD 42087, HD 148379, and HD 151804. The
comparison of mass-loss rates derived from H$\alpha$ and radio observations yielded rather good agreement,
contrary to some results from previous investigations.
However, we note that there are outliers/significant discrepancies for individual cases.

The renormalization of radio observations allowed us to derive a new calibration for the WLR for supergiants of spectral types in the range O8 to B0.

The comparison between predicted ($p$-values) and radio-derived
($r$-values) mass-loss rates yields different results for stars
above and below the BSJ temperature. For the latter group we find $p$-values
that are larger than the $r$-values, whilst we find the contrary
for the first group. The discrepancy in the first group may be
attributed to wind clumping, but this would not explain the
discrepancy for the objects below the BSJ temperature.
An investigation into the effect of wind 
clumping on the predictions, for objects both above and below the BSJ 
temperature, would shed light on this issue.\\

Our radio study of 30 supergiants of spectral types O8-B3 has
revealed that for stars around the BSJ, the wind efficiency appear
to be above the predicted general declining function towards later
spectral types. The temperature of the potential local maximum
also agrees with the predicted temperature of the BSJ. Possibly, a
similar behaviour is present for the H$\alpha$ results from SMC
stars by Trundle \& Lennon (2005). 

From the analysis of the WLR obtained here, we found the 
corresponding empirical CAK force multiplier to lie in the
range $\alpha_{\rm eff} = 0.78 - 1$. For these large values of 
$\alpha_{\rm eff}$, the concept of the WLR may loose its meaning.
 Studies of the wind efficiency as a function of 
effective temperature should be considered a powerful alternative to the 
more often used WLR.


\begin{acknowledgements}
We are indebted
to the referee, Dr. J. Cassinelli, for critical comments that
helped to improve this work,
and thank S. Johnston for assistance with observations. PB thanks
F.A. Bareilles for help in figure handling. PB was supported in
part by grant BID 1728/OC-AR PICT 03-13291 (ANPCyT) and CONICET
(PIP 5375). JSV acknowledges RCUK for his Academic Fellowship and
interesting discussions with Drs. D. Lennon and C. Trundle about B
supergiants. JM acknowledges support by grant AYA2004-07171-C02-02
of Spanish government, FEDER funds, and FQM322 of Junta de
Andaluc\'{\i}a. This work has made use of the PoP Survey, from the
UVES Paranal Observatory Project (ESO DDT Program ID 266.D-5655).

\end{acknowledgements}






\begin{thebibliography}{}

  \bibitem[]{}Abbott, D.C., Bieging, J.H., Churchwell, E., \& Cassinelli, J.P. 1980, ApJ, 238, 196

  \bibitem[]{} Bagnulo, S., Jehin, E., Ledoux, C., et al. 2003, Messenger, 114, 10

  \bibitem[]{} Benaglia, P., Cappa, C.E., \& Koribalski, B. 2001,
      A\&A, 200, 58

  \bibitem[]{} Bieging, J.H., Abbott, D.C., \& Churchwell, E.B. 1989, ApJ, 340, 518

  \bibitem[]{} Blomme, R., van de Steene, G.C., Prinja, R.K., et al. 2003, A\&A, 408, 715

  \bibitem[]{} Bohannan, B., \& Crowther, P.A. 1999, ApJ, 511, 374

  \bibitem[]{} Bouret, J.,-C., Lanz, T., \& Hillier, D.J. 2003, ApJ, 595, 1182

  \bibitem[C]{} Castor, J.I., Abbott, D.C., \& Klein, R.I.
  1975, ApJ, 195, 157

  \bibitem[]{6} Chiosi, C., \& Maeder, A. 1986, ARA\&A, 24, 329

  \bibitem[]{} Crawford, D.L., Barnes J.V., Golson J.C. 1971, AJ, 76, 1058

  \bibitem[]{} Crowther, P.A., \& Bohannan, B. 1997, A\&A,
  317, 532

  \bibitem[]{6} Crowther, P.A., Lennon, D.J., \& Walborn, N.R.
  2006, A\&A, 446, 279

  \bibitem[]{} Cure, M., Rial, D.F., \& Cidale, L. 2005, A\&A, 437, 929

  \bibitem[]{} Davies, B., Vink, J.S., \& Oudmaijer, R.D. 2007,
  A\&A, to be submitted

  \bibitem[]{} Eldridge, J.J., \& Vink, J.S., 2006, A\&A, 452, 295

  \bibitem[]{} Evans, C.J., Lennon, D.J., Walborn, N.R., et al. 2004, PASP, 116, 909

  \bibitem[]{} Eversberg, T., L\'epine, S., \& Moffat, A.F.J. 1998, ApJ, 494, 799

  \bibitem[]{} Feinstein, A., \& Marraco, H.G. 1979, AJ, 84, 1713

  \bibitem[]{} Fullerton, A.W., Massa, D.L., \& Prinja, R.K.
  2006, ApJ, 637, 1025

  \bibitem[]{} Garrison, R.F., Hiltner, W.A., \& Schild, R.E.
  1977, ApJS, 35, 111

  \bibitem[]{} Herrero, A., Puls, J., \& Najarro, F. 2002, A\&A, 396, 949

  \bibitem[]{} Heger, A., Fryer, C.L., Woosley, S.E., et al. 2003, ApJ, 591, 288

  \bibitem[]{} Hillier, D.J., \& Miller, D.L. 1998, ApJ, 496, 407

  \bibitem[]{} Hiltner, W.A., Garrison, R.F., \& Schild,
  R.E. 1969, ApJ, 1157, 313

  \bibitem[]{} Howarth, I.D., Siebert, K.W., Hussain, G.A.J., \& Prinja, R.K.
  1997, MNRAS, 284, 265


  \bibitem[]{} Humpreys, R.M. 1978, ApJS, 38, 309

  \bibitem[]{} Hunter, I., Smoker, J.V., Keenan, F.P., et al. 2006, MNRAS, 367, 1478

  \bibitem[]{} Johnson, H.L., Mitchell R.I., Iriarte B., \& Wisniewski W.Z. 1966, Comm. Lunar Plan. Lab. IV, No 63

  \bibitem[]{} Kotak, R., \& Vink, J.S. 2006, A\&A (Letters), in press

  \bibitem[]{} Kozok, J.R. 1985, AAS, 61, 387

  \bibitem[]{0} Kudritzki, R., \& Puls, J. 2000, ARAA, 629

  \bibitem[]{} Lamers, H.J.G.L.M., \& Pauldrach, A.W.A. 1991, A\&A, 244, L5

  \bibitem[]{} Lamers, H.J.G.L.M., \& Leitherer, C. 1993,
  ApJ, 412, 771

  \bibitem[]{} Lamers, H.J.G.L.M., Snow, T.P., \& Lindholm, D.M.
  1995, ApJ, 455, 269

  \bibitem[]{} Leitherer, C., Chapman, J.M., \& Koribalski, B.
  1995, ApJ, 450, 289

  \bibitem[]{} Limongi, M., \& Chieffi, A. 2006, ApJ, 647, 483

  \bibitem[]{} Lesh, J.R. 1968, ApJ, 517, 371

  \bibitem[]{} Lucy, L.B., \& Solomon, P.M. 1970, ApJ,
  159, 879

  \bibitem[]{} Maeder, A., \& Meynet, G. 2003, A\&A, 411, 543

  \bibitem[]{} Ma\'{\i}z-Apell\'aniz, J., Walborn, N.R., Galu\'e, H.A., \& Wei, L.H.
  2004, ApJS, 151, 103

  \bibitem[]{} Ma\'{\i}z Apell\'aniz, J. 2004, PASP 116, 859

  \bibitem[]{} Markova, N., Puls, J., Repolust, T., \& Markov, H. 2004, A\&A, 413, 693

  \bibitem[]{} Martins, F., Schaerer D., \& Hillier, D.J.
  2005, A\&A, 436, 1049

  \bibitem[]{} Massey, P., \& Thompson, A.B. 1991, AJ, 101, 1408

  \bibitem[]{} Moffett, A.T., \& Barnes, T.G. 1979, PASP, 91, 289

  \bibitem[]{} Mokiem, M.R., de Koter, A., Vink, J.S., et al. 2007, A\&A,
  submitted

  \bibitem[]{4} Morel, T., Marchenko, S.V., Pati, A.K., et al.
  2004, MNRAS, 351, 552

  \bibitem[]{} Morgan, W.W., Code, A.D., \& Whitford, A.E.
  1955, ApJS, 2, 41


  \bibitem[]{} Panagia, N., \& Felli, M. 1975, A\&A, 39, 1

  \bibitem[]{} Pauldrach, A.W.A., \& Puls, J. 1990, A\&A, 237, 409

  \bibitem[]{} Pelupessy, I., Lamers, H.J.G.L.M., \& Vink, J.S. 2000, A\&A, 359, 695

  \bibitem[]{} Perryman, M.A.C., Lindegren, L., Kovalevsky, J., et al. 1997, A\&A, 323, L49

  \bibitem[]{} Prinja, R.K., Barlow, K.J., \& Howarth, I.D.
  1990, ApJ, 361, 607

  \bibitem[]{} Prinja, R.K., Massa, D., \& Searle,
  S.C. 2005, A\&A, 403, L41

  \bibitem[]{} Puls, J., Markova, N., Scuderi, S., et al.
  2006, A\&A, 454, 652

  \bibitem[]{} Rauw, G., Naz\'e, Y., Carrier, F., et al. 2001, A\&A, 366, 585

  \bibitem[]{} Repolust, T., Puls, J., \& Herrero, A.
  2004, A\&A, 415, 349

  \bibitem[]{} Schild R.E., Garrison R.F., Hiltner W.A. 1983, ApJS, 51, 321

  \bibitem[]{} Scuderi, S., Bonanno, G., di Benedetto, R., et al. 1992, ApJ, 392, 201

  \bibitem[]{} Scuderi, S., Panagia, N., Stanghellini, et al. 1998, A\&A, 332, 251

  \bibitem[]{} Setia Gunawan, D.Y.A., de Bruyn, A.G., van der Hucht, K.A., \& Williams, P.M. 2000, A\&A, 356, 676

  \bibitem[]{} Smith, N., Vink, J.S., de Koter,
  A. 2004, ApJ, 615, 475

  \bibitem[]{} Thaller, M.L., Gies, D.R., Fullerton, A.W., et al. 2001, ApJ, 554, 1070

  \bibitem[]{} Trundle, C., Lennon, D.J., Puls, J., \& Dufton, P.L. 2004, A\&A, 417, 217

  \bibitem[]{} Trundle, C., \& Lennon, D.J. 2005, A\&A, 434, 677

  \bibitem[]{} Vink, J.S. 2000, Ph.D. Thesis, University of
  Utrecht

  \bibitem[]{} Vink, J.S. 2006, in ``Stellar Evolution at
  Low Metallicity: Mass Loss, Explosions, Cosmology", Eds.
  H.J.G.L.M. Lamers, N. Langer, T. Nugis, \& K. Annuk, PASP Conf. Ser., 353, 113

  \bibitem[]{} Vink, J.S., \& de Koter, A. 2002, A\&A,
  393, 543

  \bibitem[]{} Vink, J.S., de Koter, A., \& Lamers, H.J.G.L.M.
  1999, A\&A, 350, 181

  \bibitem[]{} Vink, J.S. de Koter, A., \& Lamers, H.J.G.L.M., 2000a, A\&A 362, 295

  \bibitem[]{} Vink, J.S. de Koter, A., \& Lamers, H.J.G.L.M., 2000b, ASPC, 204, 427

  \bibitem[]{}   Vink, J.S. de Koter, A., \& Lamers, H.J.G.L.M., 2001, A\&A 369, 574

  \bibitem[]{} Walborn, N.R. 1972a, AJ, 77, 312

  \bibitem[]{} Walborn, N.R. 1972b, ApJ, 176, 119

  \bibitem[]{} Walborn, N.R. 1973, AJ, 78, 1067

  \bibitem[]{} Walborn, N.R. 1976, ApJ, 205, 419

  \bibitem[]{} Walborn, N.R. 1977, ApJ, 176, 116

  \bibitem[]{} Walborn, N.R. 1982, AJ, 87, 1300

  \bibitem[]{} Williams, P.M. 1996, ASP Conf. Ser. 93, p.15

  \bibitem[]{} Wright, A.E., \& Barlow, M.J. 1975, MNRAS,
  170, 41

\end{thebibliography}
\end{document}